\documentclass[conference,a4paper]{IEEEtran}
\usepackage{bbm,dsfont,mathrsfs,fixmath,amsthm}
\usepackage{amsmath,amssymb,graphicx} 
\usepackage{stmaryrd,cite}
\usepackage{amsmath,amstext,amsfonts,amssymb}
\usepackage{epsfig}
\usepackage{exscale}
\usepackage{enumerate}
\usepackage{mathtools}
\usepackage{multirow}
\usepackage{cuted}
\usepackage{stfloats}
\usepackage[caption=false,font=footnotesize]{subfig}

\newtheorem{lemma}{Lemma}
\newtheorem{cor}{Corollary}
\newtheorem{remark}{Remark}
\newtheorem{theorem}{Theorem}

\newcommand{\cond}{\,\vert\,}

\newcommand{\bPs}{{\bar{p}_s}}
\newcommand{\bp}{{\bar{p}}}
\newcommand{\bq}{{\bar{q}}}
\newcommand{\br}{{\bar{r}}}
\newcommand{\ba}{{\bar{a}}}
\newcommand{\bkappa}{{\bar{\kappa}}}

\newfont{\bbb}{msbm10 scaled 700}

\newfont{\bb}{msbm10 scaled 1100}

\newcommand{\RR}{\mbox{\bb R}}

\newcommand{\EE}{\mbox{\bb E}}

\newcommand{\onev}{{\bf 1}}

\newcommand{\Cc}{{\cal C}}

\newcommand{\Ec}{{\cal E}}

\newcommand{\Jc}{{\cal J}}

\newcommand{\Pc}{{\cal P}}

\newcommand{\Sc}{{\cal S}}
\newcommand{\Tc}{{\cal T}}
\newcommand{\Uc}{{\cal U}}
\newcommand{\Wc}{{\cal W}}
\newcommand{\Vc}{{\cal V}}
\newcommand{\Xc}{{\cal X}}
\newcommand{\Yc}{{\cal Y}}
\newcommand{\Zc}{{\cal Z}}

\renewcommand{\arg}{{\hbox{arg}}}

\newcommand{\eqdef}{\stackrel{\Delta}{=}}

\usepackage{hyperref}
\hypersetup{
    bookmarks=true,         
    unicode=false,          
    pdftoolbar=true,        
    pdfmenubar=true,        
    pdffitwindow=false,     
    pdfstartview={FitH},    
    pdfnewwindow=true,      
    colorlinks=true,       
    linkcolor=red,          
    citecolor=green,        
    filecolor=blue,      
    urlcolor=blue           
}

\begin{document}

\title{Joint State Sensing and Communication over Memoryless Multiple Access Channels 
} 
\author{
\IEEEauthorblockN{Mari Kobayashi$^{1}$, Hassan Hamad$^{1}$, Gerhard Kramer$^{1}$, and Giuseppe Caire$^{2}$}
\IEEEauthorblockA{$^{1}$Technical University of Munich, Germany \\
$^{2}$ Technical University of Berlin, Germany\\ 
Emails: \{mari.kobayashi, hassan.hamad, gerhard.kramer\}@tum.de,~caire@tu-berlin.de }
}

\maketitle
\begin{abstract}
A memoryless state-dependent multiple access channel (MAC) is considered where two transmitters wish to convey a respective message to a receiver while simultaneously estimating the respective channel state via generalized feedback. 
The scenario is motivated by a joint radar and communication system where the radar and data applications share the same bandwidth. 
An achievable capacity-distortion tradeoff region is derived that outperforms a resource-sharing scheme through a binary erasure MAC with binary states.  
\end{abstract}

\section{Introduction}\label{section:introduction}

\begin{figure}[t]
	\begin{center}	
		\includegraphics[width=0.5\textwidth]{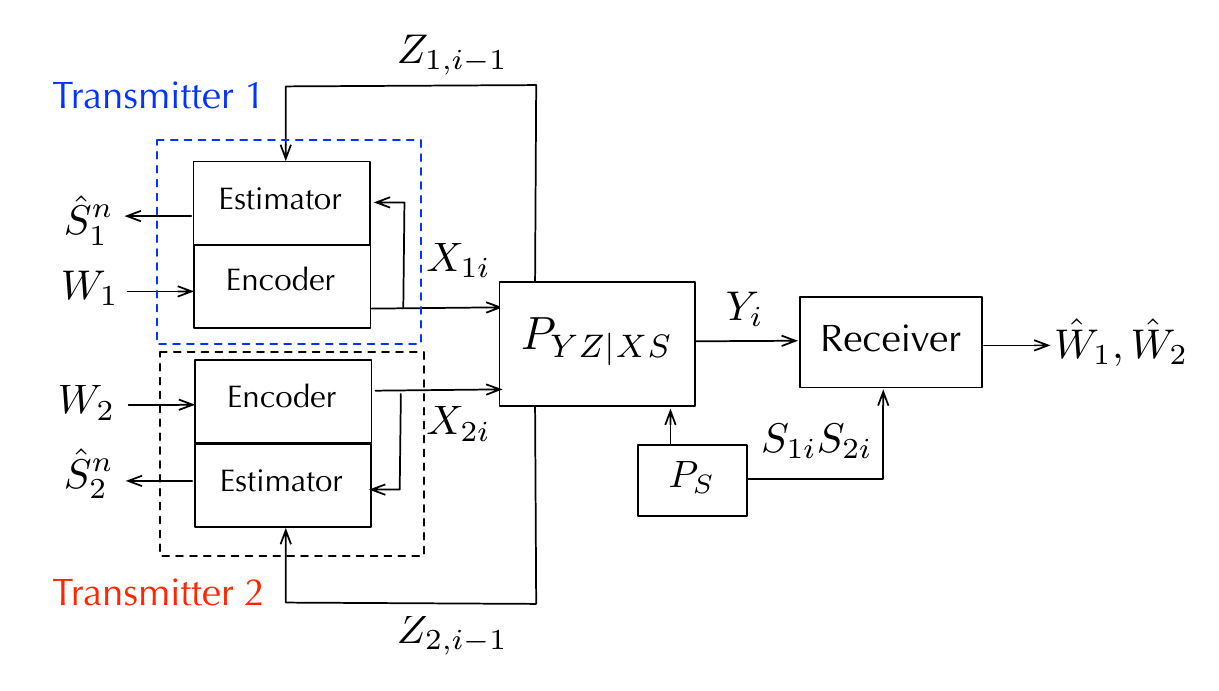}
		\caption{State-dependent MAC with generalized feedback}
	\label{fig:MAC}
	\end{center}
\end{figure}

Consider the communication setup depicted in Fig.~\ref{fig:MAC}. Two encoders each wish to 
convey a message to a decoder over a state-dependent multiple access channel (MAC) and
simultaneously estimate their state sequence via generalized feedback $Z_{k,i-1}$, $k=1,2$, $i=2,\ldots,n$.
For simplicity, we assume that at time $i$ the decoder has access to the state $S_i=(S_{1i}, S_{2i})$. 
The above communication setup is motivated by joint radar and data communications, where radar-equipped
transmitters track the state while exchanging data. Most current communication systems build on resource
sharing, where the time and frequency resources are divided into either state sensing or communication.

We recently studied a single-user version of this problem in~\cite{ISIT2018}.
In this paper, we extend the results to two-user MACs. 
As in~\cite{ISIT2018}, the state information is available at the receiver, which is
different from \cite{zhang2011joint} where the state is estimated at the receiver. 
The main contributions of the paper are:
\begin{itemize}
\item an outer bound on the capacity-distortion region that builds on \cite{tandon2011dependence};
\item an achievable rate-distortion region that builds on \cite{WillemsThesis};
\item numerical examples based on a binary erasure MAC. 
\end{itemize}

This paper is organized as follows. Section \ref{section:model} describes the model and presents our main results. Section \ref{section:converse} provides the outer bound and Section \ref{section:achievability} provides the achievability proof. We consider a binary erasure MAC with binary states in Section \ref{section:example}. 

\section{System Model and Main Results}\label{section:model}
Consider the channel inputs $X_{ki}\in \Xc_k$, the channel outputs $Y_i\in \Yc$, the feedback channel outputs $Z_{ki}\in \Zc$,
and channel state $S_i\in \Sc_1\times \Sc_2$, $k=1,2$, $i=1,\ldots,n$ linked by a discrete memoryless channel with i.i.d. states.
The joint probability distribution of these random variables can be written as
\begin{eqnarray}
& \prod_{i=1}^n P_{S}(s_{i} )  P_{Y Z_1Z_2 |X_1X_2S} (y_i, z_{1i}, z_{2i} | x_{1i}, x_{2i} ,s_i)  \nonumber \\
& \quad P(x_{1i} | x_1^{i-1}, z_1^{i-1} ) P(x_{2i} | x_2^{i-1}, z_2^{i-1}) .
\end{eqnarray}
A $(2^{nR_1},2^{nR_1}, n)$ code for the state-dependent discrete memoryless MAC with generalized feedback consists of 
\begin{itemize}
\item Two message sets $\Wc_k = [1:2^{nR_k}]$ for $k=1, 2$.
\item Encoder $k$: a function $\phi_{ki}: \Wc_k \times \Zc_k \mapsto \Xc_k$ that assigns a symbols $x_{ki}=\phi_{ki}(w_k, z_k^{i-1})$ for $i=1, \dots, n$. For simplicity, we write $x_k^n=\phi_k^n(w_k, z_k^{n-1})$ for the sequence of $n$ encoded symbols. 
 \item Decoder: a function $g: \Yc^n \times \Sc^n _1 \times \Sc^n _2 \mapsto \Wc_1\times \Wc_2$ that assigns a message pair $(\hat{w}_1, \hat{w}_2) =g(y^n, s^n)$.  
\item State estimator $k$ outputs the estimate $\hat{s}_k^n$ as a function of $x_k^n$ and $z_k^n$. We consider without loss of generality a function $\tilde{\psi}^n_k : \Xc^n _k \times \Zc^n _k\mapsto \Sc^n _k$ \cite[Lemma 2]{ISIT2018} so that $\hat{s}_k^n =\tilde{\psi}^n_k( x_k^n,z_k^n)$.   
\end{itemize}
The average distortion of estimator $k$ is
\begin{align}\label{eq:AveDistortion}
d^{(n)}_k 
&= \EE\left[ \frac{1}{n} \sum_{i=1}^n d_k(S_{ki}, \hat{S}_{ki}) \right]
\end{align}
where $d_k: \Sc_k\times \hat{\Sc}_k \mapsto [0,\infty)$ measures the distortion between a state symbol and a reconstruction symbol.
We consider bounded distortion functions with $d_{\max}\eqdef \max_{(k,s, \hat{s})} d_k(s, \hat{s})$. 
Let the average error probability be $P^{(n)}_e$. We say that $(R_1,R_2, D_1, D_2)$ is
achievable if for all $\epsilon>0$ there is some $n$ and a $(2^{nR_1},2^{nR_2}, n)$ code satisfying
$P^{(n)}_e \le \epsilon$ and  $d^{(n)}_k \le D_k+\epsilon$ for $k=1,2$.
The capacity region $\Cc(D_1, D_2)$ is the closure of achievable $(R_1,R_2)$ for
specified $D_1, D_2$. 

For our outer bound on $\Cc(D_1, D_2)$, we consider idealized transmitter
estimators $\hat{s}_k=\psi^*_k(x_1, x_2, z_1, z_2)$, $k=1,2$, that are aware of  
$x_1,x_2$ as well as $z_1, z_2$. The best such estimators are
\begin{align}\label{eq:Psi}
& \psi^*_k (x_1, x_2, z_1, z_2) = \arg \min_{\psi_k: \Xc_1\times \Xc_2 \times \Zc_1\times \Zc_2 \mapsto \Sc_k } \nonumber\\
&\sum_{s_k\in \Sc_k} P_{S_k| X_1X_2Z_1 Z_2} (s_k| x_1 x_2 z_1 z_2) d_k(s_k, \psi_k(x_1, x_2, z_1,z_2))
\end{align}
for $k=1,2$ with the conditional distortions
\begin{align}\label{eq:OptimalEst}
& c_k (x_1, x_2) = \EE[d_k(s_k, \psi^*_k(x_1, x_2, z_1, z_2)) | X_1=x_1, X_2= x_2].
\end{align}
The following outer bound extends a bound from \cite{tandon2011dependence} to state-dependent MACs
with distortion constraints.

\begin{theorem} \label{theorem:converse}
$\Cc(D_1, D_2)$ is a subset of the union of $(R_1, R_2)$ satisfying
\begin{subequations}
\begin{align}
R_1 &\leq I(X_1; Y Z_1 Z_2| S X_2 T)\label{eq:UTbound1}\\
R_2 & \leq I(X_2; Y Z_1 Z_2| S X_1 T)\label{eq:UTbound2} \\
R_1 + R_2 & \leq I(X_1 X_2; Y Z_1 Z_2| S T) \label{eq:UTbound3}\\
R_1 + R_2 &\leq I(X_1 X_2; Y|  S)\label{eq:CoopBound}
\end{align}
\end{subequations}
where $T - SX_1X_2 - YZ_1Z_2$ forms a Markov chain, and we have the dependence
balance constraint
\begin{align}\label{eq:DB-constraint}
I(X_1;X_2|T) \leq I(X_1; X_2| Z_1 Z_2 T)
\end{align}
and the average distortion constraints
\begin{align}\label{eq:Dist-constraint}
\EE[c_k (X_1, X_2)] \leq D_k, \quad k=1,2.
\end{align}
It suffices to consider $T$ whose alphabet $\mathcal{T}$ has cardinality
$|\mathcal{T}| \leq 7$ (see Appendix \ref{appendix:Cardinality}).
\end{theorem}
\begin{remark}The result yields a number of special cases studied in the literature.
Without distortion constraints and states, the bounds reduce to the ones derived in \cite{tandon2011dependence}.
For a single user, i.e., $X_2$ and $Z_2$ constants, Theorem \ref{theorem:converse} yields the
capacity-distortion tradeoff in \cite{ISIT2018}. 
For a special case when the feedback is output feedback $Z_1=Z_2=Y$ and we have no
distortion constraints, the region reduces to \cite[Section VII]{hekstra1989dependence}.
 \end{remark}

\newcommand{\uc}{{\underline{c}}}
\newcommand{\ud}{{\underline{d}}}
\newcommand{\upsi}{{\underline{\psi}}}
For our achievable region, we consider an estimator $\upsi^*_1 (x_1, v_2, z_1)$ given by 
\begin{align}\label{eq:OurPsi}
& \upsi^*_1 (x_1, v_2, z_1) = \arg \min_{\psi_1: \Xc_1\times \Vc_2 \times \Zc_1 \mapsto \Sc_1} \nonumber\\
&\sum_{s_1\in \Sc_1} P_{S_1| X_1V_2Z_k} (s_k| x_1 v_2 z_k) d_1(s_1, \psi_1(x_1, v_2, z_k))
\end{align}
yielding the estimation cost as
\begin{align}\label{eq:OurEst}
& \uc_1 (x_1, v_2) = \EE[d_1(s_1, \upsi^*_1(x_1, v_2, z_1)) | X_1=x_1 V_2= v_2]
\end{align}
We define $ \upsi^*_2 (v_1, x_2, z_2)$ and $\uc_2(v_1, x_2)$ similarly. 
The following achievable region is based on \cite{WillemsThesis}.
 
\begin{theorem}\label{theorem:achievability}
$\Cc(D_1, D_2)$ includes the $(R_1, R_2)$ satisfying
\begin{subequations}
\begin{align}
R_1 &\leq I(X_1; Y| X_2 V_1 U S) + I(V_1; Z_2| X_2 U)  \label{eq:FTbound1}\\
R_2 & \leq I(X_2; Y| X_1 V_2 U S) + I(V_2; Z_1| X_1 U) \label{eq:FTbound2} \\
R_1 + R_2 & \leq \min\{ I(X_1 X_2;Y| S), I(X_1 X_2; Y| S V_1 V_2 U)\nonumber\\
&   + I(V_1; Z_2| X_2 U) + I(V_2; Z_1| X_1 U)\} \label{eq:FTbound3}
\end{align}
\end{subequations}
where $V_1X_1 - U - V_2X_2$ and $UV_1V_2 - X_1X_2 - YZ_1Z_2$ form Markov chains,
and where
\begin{subequations}
\begin{align}\label{eq:achievable-distortion}
& \EE[\uc_1(X_1, V_2)]  \le D_1\\
& \EE[\uc_2(V_1, X_2)]  \le D_2.
\end{align}
\end{subequations}
\end{theorem}

\section{Converse}\label{section:converse}
This section provides a sketch of proof for Theorem \ref{theorem:converse}. Details are provided in Appendix \ref{appendix:converse}.
By following the same steps as \cite{tandon2011dependence} and \cite{hekstra1989dependence}, we have 
\begin{subequations} \label{eq:multiletter}
\begin{align} 
&n R_1  \le  \sum_{i=1}^n  I(X_{1i}; Y_i Z_{i}|  S_i X_{2i} Z^{i-1}) +n\epsilon \label{eq:Single-mc1} \\ 
&n R_2  \le  \sum_{i=1}^n  I(X_{2i}; Y_i Z_{i}| S_i X_{1i} Z^{i-1}) +n\epsilon \label{eq:Single-mc2} \\ 
&n(R_1 + R_2) \le \sum_{i=1}^n I(X_{1i} X_{2i}; Y_{i} Z_{i}  | S_i Z^{i-1}) +n\epsilon \label{eq:Sum-mc} \\
& \sum_{i=1}^n  I(X_{1i}; X_{2i}|Z_i Z^{i-1})  \le \sum_{i=1}^n I(X_{1i}; X_{2i}| Z^{i-1}). \label{eq:ML-DB}
\end{align}
\end{subequations}
where we let $Z_i = (Z_{1i},Z_{2i})$.
Next, suppose a genie gives both inputs $X_{1,i},X_{2,i}$ to both transmitters when estimating $S_{k,i}$ for $j\ne k$.
We then have the distortion constraints 
\begin{align}\label{eq:ML-Distortion}
\frac{1}{n} \sum_{i=1}^n 
\EE[ c_k (X_{1i}, X_{2i})] \le D_k + \epsilon, \quad k=1,2.
\end{align}
Let $Q$ be uniform over $1,2,\ldots,n$ and independent of all other random variables. Define $T= (Q, Z^{Q-1}_1,  Z^{Q-1}_2)$, $X_{1Q}= X_1$, and similarly for all other variables. 
 By letting $n\rightarrow \infty$, we readily obtain \eqref{eq:UTbound1}, \eqref{eq:UTbound2}, \eqref{eq:UTbound3}, \eqref{eq:DB-constraint} and \eqref{eq:Dist-constraint}, while \eqref{eq:CoopBound} follows from the cut set bound. %

\section{Achievability}\label{section:achievability}
We use block Markov encoding and backward decoding~ \cite{WillemsThesis}. Encoder $k$ sends $2(B-1)$ i.i.d. messages $\{w_{k1}(b), w_{k2}(b) \}_{b=1}^{B-1}$ over $n=BN$ channel uses. The messages $w_{k1}(b)\in [1, 2^{NR_{k1}}]$ and $w_{k2}(b) \in [1,2^{NR_{k2}}]$, $k=1,2$, $b=1,...,B-1$, are uniformly distributed and mutually independent. 
By letting $B\rightarrow \infty$, we obtain $R_{jk} \frac{B-1}{B}\rightarrow R_{jk}$ for any $j, k=1,2$. 
Encoder $1$'s message $w_{12}$ is decoded by encoder $2$, while encoder $2$'s message $w_{21}$ is decoded by encoder $1$ thanks to generalized feedback, yielding encoder cooperation.  
\paragraph{Codebook Generation}
 Fix a pmf $P_U(u) \prod_{k=1}^2 P_{V_k|U}(v_k|u) P_{X_k|V_k U}(x_k| v_k, u)$ and  functions $\upsi^*_1(x_1, v_2, z_1),\upsi^*_2(v_1, x_2, z_2)$ such that the distortion constraints  are satisfied. For each block $b=1, \dots, B$, we proceed as follows:
\begin{itemize}
    \item Generate $2^{N(R_{12}+R_{21})}$ sequences $u^N(j_{b-1}, k_{b-1})$, $j_{b-1}=1,...,2^{NR_{12}}$, $k_{b-1}=1,\dots,2^{NR_{21}}$,  each according to $ \prod_{i=1}^N P_U(u_i)$.    
    \item For each $(j_{b-1}, k_{b-1})$, generate $2^{NR_{12}}$ sequences $v_1^N(j_{b-1}, k_{b-1}, j'_b)$, $j'_b = 1,\dots,2^{NR_{12}}$, each according to $\prod_{i=1}^N P_{V_1|U}(v_{1i}|u_i(j_{b-1}, k_{b-1}))$. Similarly generate $v_2^N(j_{b-1}, k_{b-1}, k'_b)$, $k'_b=1,\dots,2^{NR_{21}}$.
\item For each $(j_{b-1}, k_{b-1}, j'_b)$, generate $2^{NR_{11}}$ sequences $x_1^N(j_{b-1}, k_{b-1}, j'_b, l_b)$, $l_b = 1, \ldots, 2^{NR_{11}}$, each according to $\prod_{i=1}^N P_{X_1|U V_1}(x_{1i}|u_i(j_{b-1}, k_{b-1}), v_{1i}(j'_b))$. Similarly generate $x_2^N(j_{b-1}, k_{b-1}, k'_b, m_b)$, $m_b=1,\dots,2^{NR_{22}}$.
\end{itemize}

\paragraph{Encoding}
We set $j_0=k_0=1$ and $l_B=m_B=1$. At the end of block $b$, encoder 1 finds an index $k'_b$ such that 
\begin{align}\label{eq:SucEnc1}
  \left( u^N(\cdot,\cdot) v_1^N(\cdot,\cdot, j'_b), v_2^N(\cdot,\cdot, k'_b),  x_1^N(\cdot,\cdot, j'_b, l_b), z_1^N(b)\right)\in \Tc_{\epsilon}^N
 \end{align}
 where the first two arguments of each variable are $j_{b-1}, k_{b-1}$\footnote{
If there is more than one such index, we select one of these indices uniformly at random. If there is no such index, we choose
an index from $\{1,\dots,2^{NR_{21}}\}$ uniformly at random. A similar procedure applies to decoding and shall be omitted.}. Using this estimate $k'_b$ from block $b$, encoder 1 transmits $x_1^N(j_{b}, k'_{b}, j'_{b+1}, l_{b+1})$ in block $b+1$. 
Similarly, encoder $2$ finds an index $j'_b$ such that 
\begin{align}\label{eq:SucEnc2}
\left( u^N(\cdot,\cdot),v_1^N(\cdot,\cdot, j'_b),  v_2^N(\cdot,\cdot, k'_b), x_2^N(\cdot,\cdot, k'_b, m_b), z^N_2(b) \right)\in \Tc_{\epsilon}^N.
\end{align}
Using the estimate $j'_b$ from block $b$, encoder $2$ transmits $x_2^N(j'_{b}, k_{b}, k'_{b+1}, m_{b+1})$ in block $b+1$. 
Both encoders repeat the same procedure for each $b$. 

\paragraph{Decoding}
Assuming that $(j'_b, k'_b)$ is decoded correctly in block $b+1$, 
 the decoder finds $(j_{b-1}, k_{b-1}, l_b, m_b)$ in block $b$ such that $u^N(j_{b-1}, k_{b-1})$, $v_1^N(j_{b-1}, k_{b-1}, j'_b)$, $v_2^N(j_{b-1}, k_{b-1}, k'_b)$, $x_1^N(j_{b-1}, k_{b-1}, j'_b, l_b)$, $x_2^N(j_{b-1}, k_{b-1}, k'_b, m_b)$, $s^N(b), y^N(b)$ are jointly typical. 
The decoder repeats this step for blocks $B$ to 1. 
\paragraph{State Estimation} For each block $b=1, \dots, B$, encoder $1$ puts out
\begin{align*}
\hat{s}_1^N (b) =  \upsi_1^*(x^N_1(j_{b-1}, k_{b-1}, j'_b, l_b), v^N_2(j_{b-1}, k_{b-1}, k'_b),z^N_1(b))
\end{align*}
where $k'_b$ is decoded at the end of block $b$ during encoding process. 
Similarly,  encoder 2 lets 
\begin{align*}
\hat{s}_2^N (b) = \upsi_2^*(v^N_1(j_{b-1}, k_{b-1}, j'_b), x^N_2(j_{b-1}, k_{b-1}, k'_b, m_b),z^N_1(b))
\end{align*}
where $j'_b$ is known to encoder 2 from its encoding process. 

\paragraph{Error Probability}
Following the same steps as \cite{WillemsThesis},  we can prove that by letting $N\rightarrow\infty$, $P_e^{(n)} \to 0$ if the following conditions
 hold: 
\begin{subequations}\label{eq:RateConditions}
\begin{align}
R_{12} &\leq I(V_1; Z_2| X_2 U) \\
R_{21} &\leq I(V_2; Z_1| X_1 U) \\
R_{11} &\leq I(X_1; Y| S X_2 V_1 U) \\
R_{22} &\leq I(X_2; Y| S X_1 V_2 U) \\
R_{11} + R_{22} &\leq I(X_1 X_2; Y| S V_1 V_2 U) \\
R_{12} + R_{21} + R_{11} + R_{22} &\leq I(X_1 X_2; Y|S) 
\end{align}
\end{subequations}
The analysis details are provided in Appendix \ref{appendix:ErrorProbability}.
Applying Fourier-Motzkin elimination, we obtain the desired expressions. 

\paragraph{Distortion} 
If there is no decoding error, $(u^N(b)$, $v_1^N(b)$, $v_2^N(b)$, $x_1^N(b)$, $x_2^N(b)$, $y^N(b)$, $s(b))$ are jointly typical 
for all $b$. We simplify notation and let $w_k= (w_k(1), \dots, w_k(B-1))$ with $|\Wc_k| = 2^{N(B-1) (R_{k1}+R_{k2})}$ for $k=1, 2$, where 
 $w_{k}(b)$ denotes $(w_{k1}(b), w_{k2}(b))$. 
For a given message pair $(w_1, w_2)$, we bound the average distortion for encoder 1. 
\begin{align*}
d_1^{(n)} (w_1,w_2) & \le P_e^{(n)} (w_1, w_2) d_{\max}   \\
& \quad + (1-P_e^{(n)}(w_1, w_2) )  (1+\epsilon) \EE\left[ \uc_1(X_{1}, V_{2}) \right].
\end{align*}
By averaging over all possible message pairs, we obtain the desired result. The details of the proof are provided in Appendix \ref{appendix:distortion}.

\section{Example}\label{section:example}

Consider a MAC where the state and channel inputs are binary, $S_k, X_k\in \{0,1\}$ and the channel output is ternary:
\begin{align}\label{eq:binarychannel}
   Y = S_1 X_1 + S_2 X_2.
\end{align}
Consider Hamming distance, i.e., $d(s, \hat{s}) = s \oplus \hat{s}$. For simplicity, we consider output feedback $Z_1=Z_2=Y$ and assume that $S_1$ and $S_2$ are i.i.d. Bernoulli with parameter $p_s \eqdef \Pr(S=1)$. If $p_s=1$, then this channel reduces to the binary erasure MAC with feedback, whose capacity region is the Cover-Leung region \cite{cover1981achievable,willems1982feedback} (see also \cite[Chapter 17]{el2011network}). 

We compute the optimal estimation cost. The best estimator gives either zero 
distortion or $\eta =  \min\{p_s, 1-p_s\}$ yielding the following cost for encoder 1 (see Appendix \ref{appendix:Cost}):
\begin{align}\label{eq:Cost}
c_1(0,0) &= \eta P_Y(0)   ,\;\; c_1(1,1)  = \eta P_Y(1) \nonumber\\
c_1 (0,1) &= \eta ( P_Y(0)+P_Y(1)),  \;\;\; c_1(1,0) =0
\end{align}
\subsection{Proposed Scheme} 
We characterize an achievable tradeoff between the sum rate and the symmetric distortion of our
proposed scheme. 
\begin{align}
X_k = V_k \oplus \Theta_k =U \oplus \Sigma_k \oplus \Theta_k,\;\;\; k=1,2
\end{align}
where $U, \Sigma_1, \Sigma_2, \Theta_1, \Theta_2$ are mutually independent. For the sake of simplicity, we focus on the symmetric rate $R_1=R_2$ and let 
 $U, \Sigma_k, \Theta_k$ is Bernoulli distributed with parameter $p, q, r$, respectively, for $k=1,2$. 

\paragraph{Unconstrained sum rate} We first characterize the unconstrained sum rate without distortion constraints, denoted by $R_{\rm sum-prop}(\infty)$. 
\begin{cor}\label{cor:ProposedSumRate}
The unconstrained sum rate is given by:
\begin{align}
R_{\rm sum-prop}(\infty)& =\max_{(p, q, r)} \min\{f_1(p, q, r), f_2(p, q, r)\}
\end{align}
with $f_1=f_{1a}  + 2 \{ f_{1b}  - f_{1c}\}$, where $f_{1a}, f_{1b}, f_{1c}, f_2$ are defined in \eqref{eq:f1f2} by letting $\kappa=  qr+ \bq \br$ and $\bkappa =1-\kappa$.
\end{cor}
\begin{figure*}[b]
\begin{align}
f_{1a} &= 2  \bPs p_s H_2(r)  + p_s^2H_3(r^2, 2r \br, \br^2) \nonumber\\
 f_{1b}&= -\bp \kappa[(\bPs+ p_s \kappa)\log(\bPs + p_s \kappa)  + (p_s \bkappa) \log(p_s \bkappa)]  \nonumber\\
&-  p \bkappa [( \bPs +p_s \bkappa )\log(  \bPs  +p_s \bkappa ) + (p_s \kappa) \log(p_s \kappa)]\nonumber\\
&- \bp \bkappa[(\bPs^2 + p_s\bPs \kappa)\log(\bPs^2 + p_s\bPs \kappa) +(\bPs p_s +  p_s\bPs \bkappa + p_s^2 \kappa) \log(\bPs p_s +  p_s\bPs \bkappa + p_s^2 \kappa)
+  (p_s^2 \bkappa)\log(p_s^2 \bkappa)]\nonumber\\
& - p \kappa [(\bPs^2+ p_s\bPs \bkappa)\log(\bPs^2+ p_s\bPs \bkappa)+  ( \bPs p_s+  p_s\bPs \kappa + p_s^2 \bkappa) \log( \bPs p_s+  p_s\bPs \kappa + p_s^2 \bkappa)
+ (p_s^2 \kappa)\log(p_s^2 \kappa) ]\nonumber\\
f_{1c} &=-(\bp\bq\kappa+pq\bkappa) [(\bPs+p_s \br)\log(\bPs+p_s \br)+ p_s r \log(p_s r)] \nonumber\\
&-(\bp q\kappa+p\bq\bkappa)[(\bPs+p_s r)\log(\bPs+p_s r)+ p_s \br \log(p_s \br)]\nonumber\\
&-(\bp\bq\bkappa+pq\kappa)[(\bPs^2+p_s\bPs \br)\log(\bPs^2+p_s\bPs\br)+ (p_s \bPs+ p_s\bPs r  + p_s^2 \br)\log(p_s \bPs+ p_s\bPs r  + \br)+(p_s^2 r )\log(p_s^2 r)]\nonumber\\
&-(\bp q\bkappa+p\bq\kappa)[(\bPs^2+p_s\bPs r)\log(\bPs^2+p_s\bPs r) + (p_s \bPs+ p_s\bPs \br  + p_s^2 r)\log(p_s \bPs+ p_s\bPs \br  + p_s^2 r) +(p_s^2\br )\log(p_s^2 \br)]\nonumber\\
f_2&=2 p_s\bPs H_2(p\kappa+ \bp\bkappa) + p_s^2 H_3(\bp \kappa^2 + p\bkappa^2 , 2\kappa \bkappa, p \kappa^2 +\bp \bkappa^2 )\label{eq:f1f2}
\end{align}
\end{figure*}
The proof is provided in Appendix \ref{appendix:SumRateProof}.
\begin{remark} For a special case of the erasure MAC with $p_s=1$, the functions $f_1, f_2$ simplifies into:
\begin{align*}
f_1(p, q, r)&= H_3(r^2, 2r \br, \br^2) + 2(H_2(\kappa)-H_2(r))\\
f_2(p, q, r) &= H_3(\bp \kappa^2 + p\bkappa^2 , 2\kappa \bkappa, p \kappa^2 +\bp \bkappa^2 ) 
\end{align*}
It readily follows that $f_2$ is maximized by letting $p=1/2$, yielding $H_2(2\kappa\bkappa) + \kappa^2 + \bkappa^2$. It can be proved that the sum rate is given by choosing $r=0$, yielding 
\begin{align*}
& R_{\rm sum-prop}(\infty) = \max_{q} \min\{2H_2(q), H_2(2q\bq) + q^2 + \bq^2 \}. 
\end{align*}
By choosing $q^*=0.2377$, the sum capacity of 1.5822 bit/channel use is achieved \cite{willems1982feedback}.  
\end{remark}
\paragraph{Minimum distortion} The minimum distortion $D_{\min}$ can be obtained by solving the following optimization problem.
\begin{align}\label{eq:MinDist}
 \min_{p, q,q} \sum_{(x_1, x_2)} P_{X_1, V_2} (x_1, v_2) \uc_1 (x_1, v_2) 
\end{align}
where by letting $\eta_x = \min\{ p_s x, 1-p_s x\}$ the cost function $\uc_1 (x_1, v_2) $ is given by (see Appendix \ref{appendix:Cost2}).
\begin{align} \label{eq:AchEstCost}
\uc_1(0, 0) & = \eta P_Y(0) +\eta_r P_Y(1), \;\;\; \uc_1(1,1) = P_Y(1) \eta_{\br} \nonumber\\
\uc_1(0,1) &=\eta  (P_Y(0) +P_Y(1)), \;\;\;\ \uc_1(1,0) =P_Y(1) \eta_r 
\end{align}
The solution of \eqref{eq:MinDist} is achieved by choosing $X_1=X_2=U$, yielding zero sum rate. With this choice ($q=r=0$), the 
estimation cost coincides with the idealized one. 
Intermediate points between the unconstrained sum rate and the minimum distortion can be evaluated by the parametrized
optimization similarly to the single-user case \cite{ISIT2018}.

\subsection{Resource-Sharing}
We consider a resource sharing scheme that uses feedback only for state estimation purpose. Then, we can achieve $(D_{\min}, 0)$. The other extreme point is the unconstrained sum rate point without feedback. 
After some straightforward computation, 
 we obtain:
\begin{align*}
R_{\rm sum-no-fb} (\infty)&
=\max_{P_Q P_{X_1|Q} P_{X_2|Q}} H(Y|SQ)\\
&=\max_{a}  2 p_s \overline{p}_s H_2(a) +  p_s^2  H_3(a^2, 2a\bar{a},\bar{a}^2 ) \\ 
&= 2 p_s \overline{p}_s  + \frac{3p_s^2}{2} 
\end{align*}
where the last equality holds by choosing $a=\frac{1}{2}$. 
The corresponding distortion is given by a fixed estimator independent of feedback. Namely, we consider $\hat{s}_k=0$ if $p_s<\frac{1}{2}$
and $\hat{s}_k=1$ if $p_s\geq \frac{1}{2}$. This yields the distortion of $\eta=\min\{p_s, 1-p_s\}$.  In summary, the resource sharing scheme achieves any tradeoff between $(D_{\min}, 0)$ and $(\eta, R_{\rm sum-no-fb})$.

\subsection{Outer Bound} By applying the upper bound \eqref{theorem:converse} to the binary erasure MAC with binary states, we have
\begin{subequations}
\begin{align}
R_k &\leq H(X_k| S X_j T), \; \forall k=1, 2, \forall j\neq k\\
R_1 +R_2 &\leq H(Y|ST) \leq H(Y)
\end{align}
\end{subequations}
We apply the technique used in \cite{willems1982feedback} to the state-dependent erasure MAC. By focusing on the symmetric rate, we define $p_t \eqdef \Pr(T=t)$, $a_t \eqdef \Pr(X_k=1|T=t)$ for $k=1,2$.
By noticing that $H(Y|(s_1, s_2), X_2 T)$ is positive only for $(s_1, s_2)=(1,0), (11)$ and $H(Y|(s_1, s_2), X_1 T)$ is positive only for $(s_1, s_2)=(01), (11)$, it readily follows that 
\begin{align}\label{eq:807}
 H(Y|S X_2 T) &= H(Y|S X_1 T)  = p_s \sum_t p_t H_2(a_t) \nonumber \\
 &= p_s H_2(\phi(2 \sum_t p_t  a_t \bar{a}_t)) 
 \end{align}
 where we defined a function $\phi(t)=\frac{1}{2} (1-\sqrt{1-2t})$ for $t\in [0,1/2]$ and used the concavity of $H_2(\phi(t))$. 
 We also have
 \begin{align}\label{eq:811}
H(Y)&= H_3(\bPs^2+ 2 p_s\bPs\sum_t p_t \ba_t+ p_s^2 \sum_t p_t \ba_t^2, \nonumber\\
&\;\;\;\; 2 p_s\bPs\sum_t p_t a_t+2 p_s^2 \sum_t p_t a_t\ba_t, p_s^2 \sum_t p_t a_t^2 ) \nonumber\\
&\leq H_2\left(2p_s\bPs\sum_t p_t a_t+2 p_s^2 \sum_t p_t  a_t\ba_t \right)\nonumber\\
& 1-[2 p_s\bPs\sum_t p_t a_t+p_s^2 \sum_t p_t (2 a_t\ba_t )]
\end{align}
where the last inequality follows from $H_3(a, b, c)=\frac{H_3(a, b, c)+H_3(c, b, a)}{2}  \leq  H_2(b)+1-b$, By noticing that the bounds in \eqref{eq:807} and \eqref{eq:811} depend only on two parameters $\alpha=2\sum_t p_t a_t \ba_t$ and $\gamma=\sum_t p_t a_t$, we readily obtain 
\begin{align*}
R_{\rm sum-out}(\infty) &=\max_{\alpha, \gamma}\min\{2 p_s H_2(\phi(\alpha)) , H_2(2p_s \bPs \gamma + p_s^2 \alpha )\nonumber\\
&\;\;\;\;\;\; + 1-(2p_s \bPs \gamma + p_s^2 \alpha)\}  \\
&= \max_{\beta,\gamma} \min\{2 p_s H_2(\beta),H_2(2p_s \bPs \gamma + 2 p_s^2 \beta \bar{\beta} )\nonumber\\
 &\;\;\;\;\;\; + 1-(2p_s \bPs \gamma + 2 p_s^2  \beta \bar{\beta})\}. 
\end{align*}
where the last equality follows by letting $\beta=\phi(\alpha)$, or equivalently $\alpha=2\beta\bar{\beta}$. 
The minimum distortion can be calculated similarly to \eqref{eq:MinDist} by replacing the estimation cost $\uc_1(x_1,v_2)$ with 
the idealized estimation cost $c_1(x_1, x_2)$.


\subsection{Numerical Result}
\begin{figure}[t]
	\begin{center}	
		\includegraphics[width=0.5\textwidth]{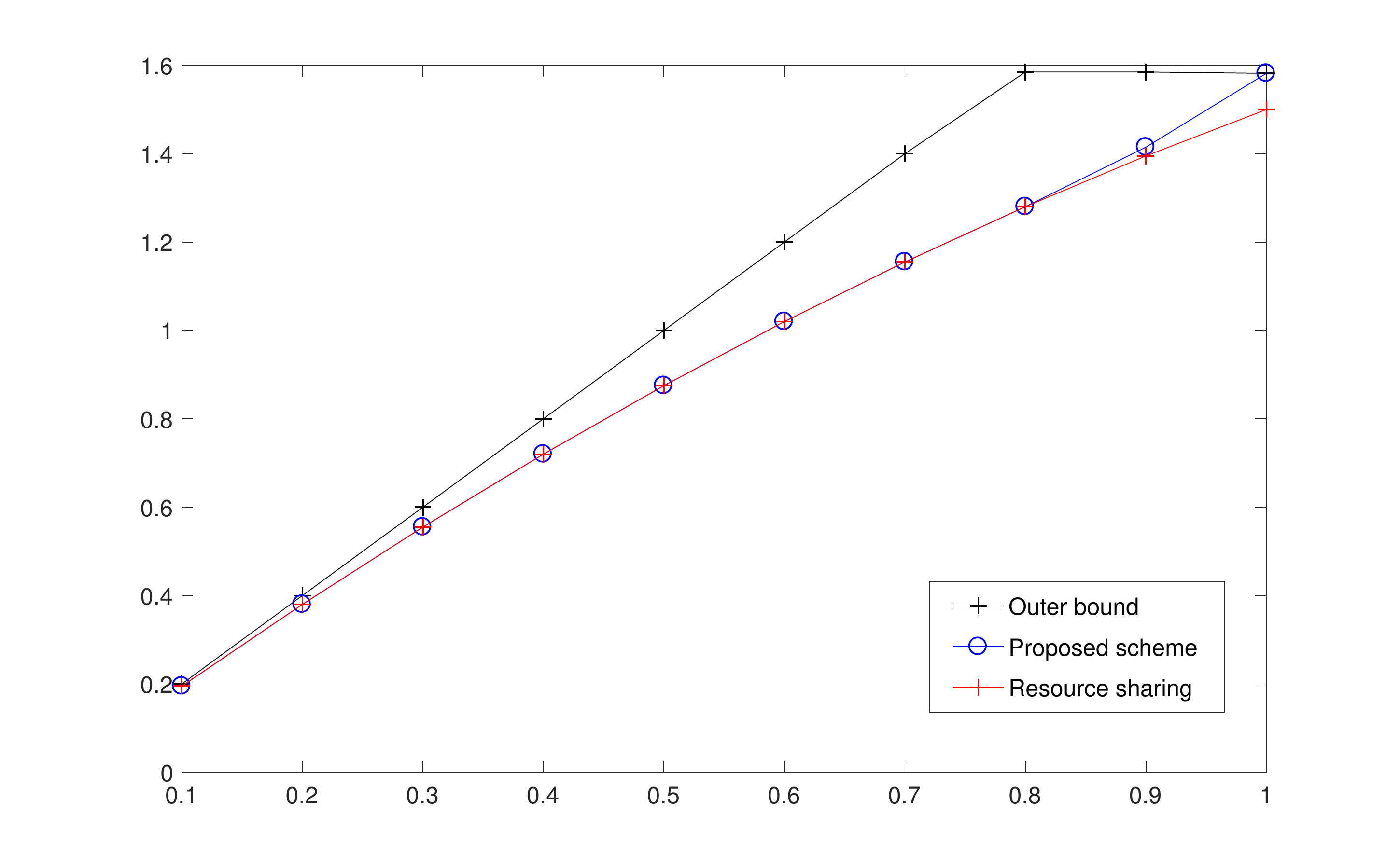}
		\caption{Unconstrained sum rate vs. state probability $p_s$.}
	\label{fig:sumrate}
	\end{center}
\end{figure}
\begin{figure}[t!]
	\begin{center}	
		\includegraphics[width=0.5\textwidth]{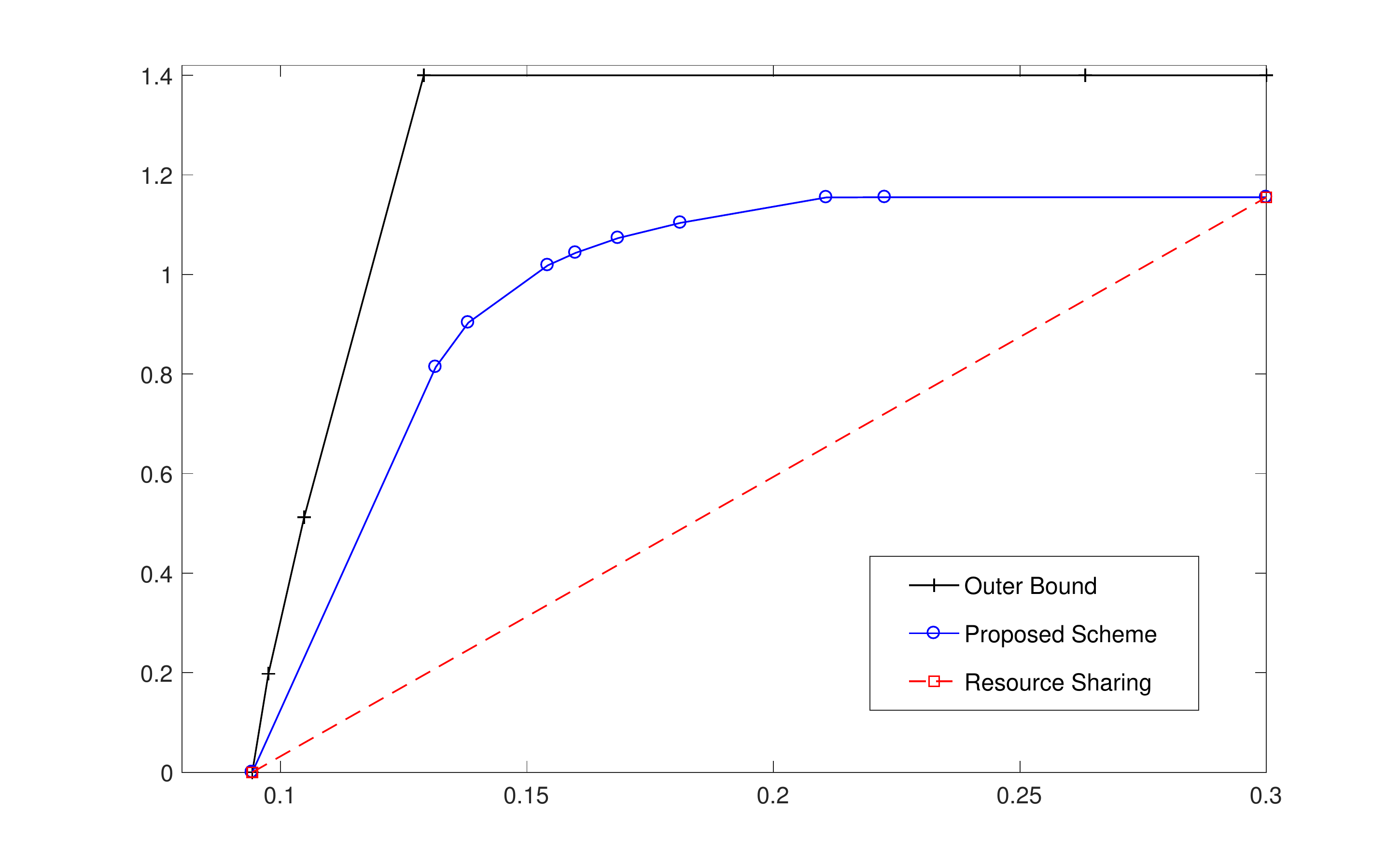}    
		\caption{Tradeoff between sum rate and distortion for $p_s=0.7$.}
	\label{fig:tradeoff}
	\end{center}
\end{figure}

Fig. ~\ref{fig:sumrate} shows the unconstrained sum rate performance as a function of the state probability $p_s$. For the case of $p_s=1$, the sum capacity is 1.5822 bit/channel use. The proposed scheme yields a visible gain with respect to the resource-sharing for $p_s>0.8$ when feedback becomes useful for the unconstrained sum rate. 
The outer bound is not very tight for $p_s$ closed to one. Fig.~\ref{fig:tradeoff} shows the tradeoff between the sum rate and the symmetric distortion for $p_s=0.7$. The proposed scheme
achieves a significant gain compared to the resource sharing scheme in terms of tradeoff. Moreover, the proposed scheme achieves near-optimal performance for small distortion values. 

Although restricted to a very simple setup, the current work demonstrates a high potential of joint sensing and communication, that exploits feedback both 
for state sensing and communication. 

\bibliographystyle{IEEEbib}

\newpage
\appendices
\section{Proof of Theorem \ref{theorem:converse}}\label{appendix:converse}
First we derive single-user bounds. As above, we define $Z = (Z_1, Z_2)\in \Zc$, and bound
\begin{align} \label{eq:R1-mc}
n R_1 & = H(W_1) = H(W_1 | W_2)  \nonumber \\
&=  I(W_1; Y^n S^n Z^n  | W_2) + H(W_1| W_2 Y^n S^n Z^n) \nonumber  \\
&\stackrel{(a)}\leq  I(W_1; Y^n S^n Z^n |W_2) + n \epsilon  \nonumber\\
& = \sum_{i=1}^n I(W_1; Y_i S_i Z_{i} | W_2 Y^{i-1} S^{i-1} Z^{i-1}) + n \epsilon  \nonumber \\
&= \sum_{i=1}^n H(Y_i S_i Z_{i} | W_2 Y^{i-1} S^{i-1} Z^{i-1}) \nonumber\\
   &~~ - H(Y_i S_i Z_{i} | W_1 W_2 Y^{i-1} S^{i-1} Z^{i-1}) +n\epsilon  \nonumber \\
&\stackrel{(b)}= \sum_{i=1}^n H(Y_i S_i Z_{i} |W_2 X_{2i} Y^{i-1} S^{i-1} Z^{i-1}) \nonumber\\
& ~~- H(Y_i S_i Z_{i} |W_1 W_2 X_{1i} X_{2i} Y^{i-1} S^{i-1} Z^{i-1}) +n\epsilon  \nonumber\\
&\stackrel{(c)} \leq  \sum_{i=1}^n H(Y_i S_i Z_{i} | X_{2i} Z^{i-1}) \nonumber\\
& - H(Y_i S_i Z_{i} |W_1 W_2 X_{1i} X_{2i} Y^{i-1} S^{i-1} Z^{i-1}) +n\epsilon  \nonumber\\
&\stackrel{(d)} =  \sum_{i=1}^n H(Y_i S_i Z_{i} | X_{2i} Z^{i-1}) \nonumber\\
&- H(Y_i S_i Z_{i} | X_{1i} X_{2i} Z^{i-1}) +n\epsilon  \nonumber\\
& = \sum_{i=1}^n  I(X_{1i}; Y_i S_i Z_{i} | X_{2i} Z^{i-1}) +n\epsilon 
\end{align}
where (a) follows from Fano's inequality; (b) follows by applying the encoding function in both terms; (c) follows by removing conditioning on $W_2,  Y^{i-1}, S^{i-1}$ in the first term;  (d) follows from the Markov chain $(W_1, W_2,Y^{i-1}, S^{i-1}) - (X_{1i}, X_{2i}, Z^{i-1}) - (Y_i, S_i, Z_{i})$. Following similar steps, we obtain also 
\begin{align}
n R_2 & \leq \sum_{i=1}^n  I(X_{2i}; Y_i S_i Z_{i} | X_{1i} Z^{i-1}) +n\epsilon  \label{eq:R2-mc}\\  
n(R_1 + R_2) &\leq \sum_{i=1}^n I(X_{1i} X_{2i}; Y_{i} S_i Z_{i}  | Z^{i-1}) +n\epsilon \label{eq:Rsum-mc}
\end{align}
The proof of the dependence balance constraint follows the same steps as \cite{hekstra1989dependence}. We start from 
\begin{align} 
0 &\leq I(W_1; W_2 | Z^n) \nonumber \\
&\stackrel{(a)}=  I(W_1; W_2 | Z^n)  - I(W_1;W_2)  \nonumber\\
&\stackrel{(b)}= -I_3(W_1; W_2; Z^n)  \nonumber \\
 &= - \sum_{i=1}^n  I_3(W_1; W_2; Z_i|Z^{i-1})  \nonumber \\
 &\stackrel{(c)} =- \sum_{i=1}^n [ H(Z_i|Z^{i-1} ) - H(Z_i |W_1 Z^{i-1}) \nonumber\\
 & - H(Z_i| W_2 Z^{i-1}) + H(Z_i| W_1 W_2 Z^{i-1})]  \nonumber \\
&\stackrel{(d)} =- \sum_{i=1}^n [  H(Z_i|Z^{i-1} ) - H(Z_i |X_{1i} W_1 Z^{i-1}) \nonumber\\
&- H(Z_i| X_{2i} W_2 Z^{i-1} ) + H(Z_i| X_{1i} X_{2i} W_1 W_2 Z^{i-1}) ]  \nonumber\\
&\stackrel{(e)} =- \sum_{i=1}^n [  H(Z_i|Z^{i-1} ) - H(Z_i |X_{1i} ) \nonumber \\
&- H(Z_i| X_{2i} ) + H(Z_i| X_{1i} X_{2i} )]  \nonumber\\
& = \sum_{i=1}^n - I_3(X_{1i}; X_{2i}; Z_i | Z^{i-1})   \nonumber \\
&\stackrel{(f)}= \sum_{i=1}^n I(X_{1i}; X_{2i}|Z_i Z^{i-1}) - I(X_{1i}; X_{2i}| Z^{i-1}) \label{eq:45}
\end{align}
where (a) follows because $W_1$ and $W_2$ are independent; (b) follows from the definition $I_3(A;B;C) = I(A;B) - I(A;B|C)$; (c) follows from the definition $I_3(A;B;C)= H(A)+ H(B)+H(C) - H(AB)-H(AC)-H(BC) + H(ABC)$; (d) follows by applying the encoding functions in last three terms; (e) follows from the Markov chain $(W_1, W_2, Z^{i-1}) - (X_{1i}, X_{2i}) - Z_i$; (f) follows from the same recursive expression used in (b). 

Now, we study the distortion constraints. From the definition, a $(2^{nR_1}, 2^{nR_2}, n)$ code must satisfy for $k=1,2$
\begin{align}\label{eq:def}
&\frac{1}{n |\Wc_1||\Wc_2|} \sum_{(w_1, w_2) }\sum_{i=1}^n  \nonumber \\
&\EE\left[ \EE[d_k(S_{k,i}, \hat{S}_{k,i}) \cond X^n_k = \phi^n_k(w_k, Z_k^{n}), \forall k] \right] \leq D_k +  \epsilon_n
\end{align}
where the outer expectation is w.r.t. $Z^n$ while the inner expectation is w.r.t. $S^n_k$ conditioned on $Z^n, w_1, w_2$. 
Notice that $\hat{S}_{k,i}$ is a deterministic function of $X_{k}^n, Z_k^n$. 
Assuming that a genie provides the other encoder's input $X_{ji}$ and $Z_j^n$ to encoder $k$ when estimating $S_{ki}$ for $j\neq k$, 
the LHS of \eqref{eq:def} can be written as: 
\begin{align}
& \frac{1}{n |\Wc_1||\Wc_2|} \sum_{(w_1, w_2)}  \sum_{i=1}^n \nonumber\\
& \min_{{\psi}'_k : \Xc _1\times \Xc_2 \times  \Zc^n\mapsto \Sc_k} \EE[ d_k(S_{k,i}, {\psi}'_k(X_{1i},X_{2i},Z^n) ) \cond X_{1i} X_{2i} ]  
\end{align}
where the expectation is w.r.t. the joint distribution of $S_k^n$, $Z^n$ conditioned on $X_{1i}, X_{2i}$. In order to proceed further, we use the following useful lemma. 
 \begin{lemma} \cite[Lemma 1]{zhang2011joint}
 For three arbitrary random variables $U\in \Uc, V\in \Vc, J \in \Jc$, satisfying the Markov chain $U- V- J$, and for an arbitrary function $d: \Uc\times \Uc\mapsto \RR$, we have
 \begin{align}
 \min_{f: \Vc\mapsto \Uc} \EE[d(U, f(V))] = \min_{g: \Vc\times \Jc \mapsto \Uc}  \EE[d(U, g(V,J))] 
 \end{align}
 \end{lemma} 
We apply this lemma by letting $U=S_{ki}, V=(X_{1i},X_{2i}, Z_{1i}, Z_{2i})$, and $J=( \{Z_{l}\}_{l\neq i})$ and noticing 
\[ S_{ki}  -X_{1i},X_{2i}, Z_{i} - \{Z_{l}\}_{l\neq i},  \] 
forms a Markov chain, we have 
\begin{align}
&\min_{{\psi}'_k : \Xc _1\times \Xc_2 \times \Zc^n _1 \Zc^n \mapsto \Sc_k}
\EE\left[  d_k(S_{k,i}, {\psi}'_k(X_{1i},X_{2i}, Z^n) ) \cond X_{1i}X_{2i}  \right] \nonumber\\
&= 
\min_{\psi_k: \Xc_1\times\Xc_2\times \Zc_k \mapsto \Sc_k}\EE\left[ d_k(S_{ki}, \psi_k(X_{1i}, X_{2i},Z_{i}) \cond X_{1i}X_{2i}   \right] \nonumber\\
&= c_k (x_{1i}, x_{2i})
\end{align}
where the last equality is from the definition \eqref{eq:OptimalEst}.
This further simplifies the distortion constraints into for $k=1,2$
\begin{eqnarray}\label{eq:563}
\frac{1}{n |\Wc_1||\Wc_2|} \sum_{(w_1, w_2) \in \Wc_1\times \Wc_2}  \sum_{i=1}^n
 c_k (x_{1i}, x_{2i})  \leq D_k +  \epsilon_n,
\end{eqnarray}
Notice that the empirical input distribution $P_{X_{1i}, X_{2i}}(x_1, x_2)$ is induced by the uniformly selected message pair, i.e. $(X_{1i}(w_1), X_{2i}(w_2))$ with probability $\frac{1}{|\Wc_1|\Wc_2|}$ for every pair $(w_1, w_2)\in \Wc_1\times \Wc_2$. Hence,  
for a sufficiently large $n$, the conditions \eqref{eq:563} reduce to for $k=1,2$
\begin{align}\label{eq:ML-Distortion}
\frac{1}{n} \sum_{i=1}^n \sum_{(x_1, x_2)\in \Xc_1\times \Xc_2}P_{X_{1i}X_{2i}} (x_1, x_2) c_k (x_1, x_2)\leq D_k + \epsilon_n 
\end{align}

{\bf Combining bounds}\\
We will combine the multi-letter upper bounds \eqref{eq:R1-mc}, \eqref{eq:R2-mc} and \eqref{eq:Rsum-mc} together with the multi-letter constraints \eqref{eq:45} and \eqref{eq:ML-Distortion}. To this end, we introduce a uniformly distributed random variable $Q \in [1, n]$ independent of all other variables as well as an auxiliary random variable $T= (Q, Z^{Q-1})$. By letting $X_{1Q}= X_1,X_{2Q}= X_2, Y_{Q}= Y, Z_{Q}=Z$ and letting $n\rightarrow \infty$, we readily obtain
\begin{align}
R_1  & \leq I(X_1; Y S Z | X_2 T) \\
R_2  & \leq I(X_2; Y S Z | X_1 T) \\
R_1 + R_2 &\leq I(X_1 X_2; Y S Z| T)\\
0 &  \leq I(X_1; X_2 |Z T) - I(X_1; X_2| T). 
\end{align}
\eqref{eq:CoopBound} follows from the cut set bound
\[ R_1 + R_2 \leq I(X_1 X_2; Y S) = I(X_1 X_2; Y |S). \]
As $n\rightarrow \infty$, we have also 
\begin{align}
\EE[ c_k (x_1, x_2)]\leq D_k.
\end{align}
This establishes the converse proof. 
\section{Proof of Cardinality Constraint on $T$}\label{appendix:Cardinality}
First we remark that the dependence balance constraint \eqref{eq:DB-constraint}, i.e. $I(X_1;X_2|T) \leq I(X_1; X_2| Z T)$, can be rewritten as
\begin{align}
I(X_1 X_2; Z| T)  \leq I(X_1;  Z|  X_2 T)+I(X_2;  Z| X_1 T).
\end{align}
We can see this easily 
\begin{align*}
I(X_1 X_2; Z| T) & =I(X_1 ; Z|X_2 T) + I(X_2; Z|T) \\
& \leq I(X_1 ; Z|X_2 T) + I(X_2;  Z| X_1 T)
\end{align*}
where the last inequality follows because 
\begin{align*}
  I(X_2;  Z| X_1 T) &= I(X_2; X_1 Z |T) - I(X_1; X_2|T) \\
  &\geq  I(X_2; X_1 Z |T)  - I(X_2; X_1| Z T) \\
  &= I(X_2;Z|T).
  \end{align*}
Let $\Pc_k$ be a subset of pmfs on $\Xc_k$ and let 
$p(x_1,x_2|t) \in \Pc\eqdef \Pc_1\times \Pc_2$, indexed by $t\in \Tc$ for an arbitrary set $\Tc$, 
be a collection of conditional pmfs on $\Xc_1 \times\Xc_2$. 
Consider the following functions that map an element of $\mathcal{P}$ into an element of $\RR$
\begin{subequations}
	\begin{align}
	g_1(P_{X_1,X_2}) & \eqdef I(X_1; Y Z| SX_2 )     \label{Cbound2} \\
	g_2(P_{X_1,X_2}) & \eqdef I(X_2; Y Z| SX_1 )     \label{Cbound3}\\
	g_3(P_{X_1,X_2}) & \eqdef  I(X_1 X_2; Y Z|S )      \label{Cbound4}\\
	g_4(P_{X_1,X_2}) & \eqdef I(X_1 X_2; Y|S)      \label{Cbound5} \\
	g_5(P_{X_1,X_2}) & \eqdef I(X_1 X_2 ; Z ) \\
	g_6(P_{X_1,X_2}) & \eqdef   I(X_1 ; Z | X_2 )  \\
	g_7(P_{X_1,X_2}) & \eqdef I(X_2 ; Z | X_1 ) 
	\end{align}
\end{subequations}
 By using  the support Lemma \cite{el2011network}, we find
\begin{align*}
I(X_1; YZ| S X_2 T) &= \int_{\mathcal{T}} g_1(p(x_1,x_2|t)) dF(t) \\ 
& = \int_{\mathcal{T}} I(X_1; YZ | S X_2 T =t) dF(t) \\
& = \sum_{t'} I(X_1; YZ| S X_2  T' =t') p(t') 
\end{align*}
Similarly, we can express all other mutual informations in terms of $T'$.
Now we have shown that considering only random variables $T'$ with constraint $|\mathcal{T'}| \leq 7$ preserves all the quantities in our outer bound. This establishes the proof of the cardinality bound. 
\section{Analysis of Error Probability}\label{appendix:ErrorProbability}

\begin{figure*}[b]
\begin{small}
\begin{equation}\begin{aligned}
& \Ec_1(k'_b) =\left\{ ( u^N(j_{b-1}, k_{b-1}), v_1^N(j_{b-1}, k_{b-1}, j'_b), v_2^N(j_{b-1}, k_{b-1}, k'_b),  x_1^N(j_{b-1}, k_{b-1}, j'_b, l_b), z_1^N(b))\in \Tc_{\epsilon}^N \right \}\\
& \Ec_2(j'_b) = \left\{ ( u^N(j_{b-1}, k_{b-1}), v_1^N(j_{b-1}, k_{b-1}, j'_b),  v_2^N(j_{b-1}, k_{b-1}, k'_b), x_2^N(j_{b-1}, k_{b-1}, k'_b, m_b), z^N_2(b))\in \Tc_{\epsilon}^N \right\}\\
&\Ec_3(j_{B-1}, k_{B-1}) \\
&= \left\{( u^N(j_{B-1}, k_{B-1}), v_1^N(j_{B-1}, k_{B-1}, 1), v_2^N(j_{B-1}, k_{B-1}, 1), x_1^N(j_{B-1}, k_{B-1}, 1,1),x_2^N(j_{B-1}, k_{B-1}, 1, 1), 
\tilde{y}^N(B)) \in \Tc_{\epsilon}^N \right\} \\
&\Ec_4(j_{b-1}, k_{b-1}, l_b, m_b) \\
&= \{( u^N(j_{b-1}, k_{b-1}), v_1^N(j_{b-1}, k_{b-1}, j'_b), v_2^N(j_{b-1}, k_{b-1}, k'_b), x_1^N(j_{b-1}, k_{b-1}, j'_b, l_b),
x_2^N(j_{b-1}, k_{b-1}, k'_b, m_b),  \tilde{y}^N(b)) \in \Tc_{\epsilon}^N\}\\
&\Ec_5(l_1, m_1) = \{( u^N(1,1), v_1^N(1, 1, j'_b), v_2^N(1, 1, k'_b), x_1^N(1,1, j'_b, l_b),x_2^N(1,1, k'_b, m_b), \tilde{y}^N(b)) \in \Tc_{\epsilon}^N\}\label{eq:events}
\end{aligned}
\end{equation}
\end{small}
\end{figure*}

We define the following events in \eqref{eq:events}. $\Ec_1(k'_b)$ and $\Ec_2(j'_b)$ are the events corresponding to \eqref{eq:SucEnc1} and \eqref{eq:SucEnc2} such that encoders 1 and 2 find a jointly typical index $k'_b, j'_b$ at the end of block $b$, respectively. 
Next, we consider the decoding events separately for $b=B$, $b=B,\dots, 2$, and $b=1$.
By recalling that there are no more fresh messages to send in block $B$, i.e. $j'_B=l_B=1$ and $k'_B=m_B=1$, we let $\Ec_3(j_{B-1}, k_{B-1})$ denote the event  that the decoder finds jointly typical indices $(j_{B-1}, k_{B-1})$, where we let $\tilde{y} = (y, s)$ denote the augmented channel output including the state. 
For $b=2, \dots, B$, assuming that the decoding of $(j'_b, k'_b)$ is done successfully in block $b+1$, we define the event $\Ec_4(j_{b-1}, k_{b-1}, l_b, m_b)$. 
Finally, for $b=1$ since we have $j_{0}= k_{0}=1$, we define the event $\Ec_5(l_1, m_1)$. 
We can now bound the average error probability as

\begin{small}
\begin{align}
P_e^{(n)} &\leq 
\Pr \Bigg\{ \bigcup_{b=1}^{B-1}\Big( \overline{\Ec}_1(k'_b) \bigcup (\cup_{k \neq k'_b} \Ec_1(k) \Big) \nonumber\\
&   \bigcup \bigcup_{b=1}^{B-1} \Big( \overline{\Ec}_2(j'_b)  \bigcup (\cup_{j \neq j'_b} \Ec_2(j))  \Big) \nonumber\\
& \bigcup  \overline{\Ec}_3(j_{B-1}, k_{B-1})  \bigcup (\cup_{(j, k) \neq (j_{B-1}, k_{B-1}) } \Ec_3(j, k) )  \nonumber\\
& \bigcup \bigcup_{b=2}^{B-1} \Big(  \overline{\Ec}_4(j_{b-1}, k_{b-1}, l_b, m_b) \nonumber\\
&\bigcup (\cup_{\substack{ (j, k, l, m) \neq \\ (j_{b-1}, k_{b-1}, l_b, m_b)}} \Ec_4(j, k, l, m) )\Big)  \nonumber\\ 
& \bigcup \overline{\Ec}_5(l_1, m_1)\bigcup (\cup_{(l, m) \neq (l_1, m_1)}  \Ec_5(l, m))  \Bigg\}
\end{align}
\end{small}
By considering all possible error events and then applying the union bound, we obtain 
\begin{small}
\begin{align}\label{eq:UnionBound}
P_e^{(n)} &\leq  
 \sum_{b=1}^{B-1} \Pr\{\overline{\Ec}_1(k'_b) \}  + \sum_{b=1}^{B-1} \sum_{k \neq k'_b} \Pr\{ \Ec_1(k)\} +  \sum_{b=1}^{B-1} \Pr\{ \overline{\Ec}_2(j'_b) \}  \nonumber\\ 
&  + \sum_{b=1}^{B-1} \sum_{j \neq j'_b} \Pr\{\Ec_2(j)) \} + \Pr\{ \overline{\Ec}_3(j_{B-1}, k_{B-1})  \} \nonumber\\ 
& + \sum_{\substack{ (j, k) \neq \\(j_{B-1}, k_{B-1})}} \Pr\{\Ec_3(j, k)\}+ \sum_{b=2}^{B-1} \Pr\{  \overline{\Ec}_4(j_{b-1}, k_{b-1}, l_b, m_b)  \}  \nonumber\\
& + \sum_{b=2}^{B-1}\sum_{\substack{ (j, k, l, m) \neq \\(j_{b-1}, k_{b-1}, l_b, m_b)}}\Pr\{\Ec_4(j, k, l, m)\}  \nonumber\\ 
& + \Pr\{ \overline{\Ec}_5(l_1, m_1) \} + \sum_{(l, m) \neq (l_1, m_1)} \Pr\{\Ec_5(l, m)\} 
\end{align}
\end{small}
In order to further simplify the upper bound on the error probability, we assume without loss of generality that $k'_b=j'_b=1$ for $b=1, \dots, B-1$,  $j_{b-1}=k_{b-1}=1$ for $b=2,\dots, B$, and $l_b=m_b=1$ for $b=1, \dots, B-1$. 
\begin{small}
\begin{align}\label{eq:UnionBound2}
P_e^{(n)} &\leq  
 (B-1) \Pr\{\overline{\Ec}_1(1) \}  + (B-1)\sum_{k \neq 1} \Pr\{ \Ec_1(k)\} \\ \nonumber
& +  (B-1) \Pr\{ \overline{\Ec}_2(1) \}  + (B-1) \sum_{j \neq 1} \Pr\{\Ec_2(j)) \} \\ \nonumber
& + \Pr\{ \overline{\Ec}_3(1,1)  \} + \sum_{ (j, k) \neq (1,1)} \Pr\{\Ec_3(j, k)\}  \\ \nonumber
& + (B-2) \Pr\{  \overline{\Ec}_4(1,1,1,1)  \} + (B-2) \sum_{\substack{(j, k, l, m)\\ \neq (1,1,1,1)}}  
\Pr\{ \Ec_4(j, k, l, m)  \}  \\ \nonumber
& + \Pr\{ \overline{\Ec}_5(1,1) \} + \sum_{(l, m) \neq (1,1)} \Pr\{\Ec_5(l, m)\} 
\end{align}
\end{small}
where we can further express 
\begin{small}
\begin{align*}
& \sum_{(j, k, l, m) \neq (1,1,1,1)}  \Pr\{ \Ec_4(j, k, l, m) \}  \\
&=  \sum_{(j,k)\neq (1,1)}\sum_{(l, m)} \Pr\{ \Ec_4(j, k, l, m)  \} + \sum_{l\neq 1, m\neq 1}\Pr\{ \Ec_4(1, 1, l, m)  \}\nonumber\\
& + \sum_{l\neq 1}  \Pr\{ \Ec_4(1, 1, l, 1)  \} +  \sum_{ m\neq 1}\Pr\{ \Ec_4(1, 1,1, m)  \}
\end{align*}
\end{small}

By the law of large numbers, as $N \to \infty$, we obtain $ \Pr\{\overline{\Ec}_1(1) \} \rightarrow 0$. The same holds for $\Pr\{ \overline{\Ec}_2(1) \}$, $\Pr\{ \overline{\Ec}_3(1,1)\}$, 
$\Pr\{  \overline{\Ec}_4(1,1,1,1)  \}$, and $\Pr\{ \overline{\Ec}_5(1,1) \}$. 
We examine the remaining error probabilities. We obtain for $k\neq 1$

\begin{small}
\begin{align}
   & \Pr\{ \Ec_1(k)  \} \nonumber\\
   &= \sum_{\substack{ (u^N, v_1^N,  v_2^N,\\ x_1^N, z_1^N) \in \Tc_{\epsilon}^N} } P_{U V_1 V_2  X_1 Z_1} (u^N v_1^Nv_2^N x_1^N  z_1^N) \nonumber \\
   &\stackrel{(a)}=  \sum_{\substack{ (u^N, v_1^N,  v_2^N,\\ x_1^N, z_1^N) \in \Tc_{\epsilon}^N} } P_{U V_1 V_2 X_1} (u^N v_1^N v_2^N x_1^N) P_{Z_1|U V_1X_1}( z_1^N | u^Nv_1^N x_1^N)\nonumber\\
   &\stackrel{(b)} \leq 2^{N(H(UV_1 V_2 X_1 Z_1) +\delta)} 2^{-N(H(U V_1 V_2 X_1)-\delta)} 2^{-N(H(Z_1|U V_1X_1)-\delta)} \nonumber\\
   &\leq 2^{-N ( I(V_2;Z_1|U V_1 X_1)-\delta)}  \nonumber\\
   &\stackrel{(c)}=  2^{-N ( I(V_2;Z_1|U X_1)-\delta)} \label{eq:E1}
 \end{align}
\end{small} 
where (a) follows from the Markov chain $V_2- U V_1 X_1 -Z_1$ for $k\neq 1$; (b) follows by noticing that the typical set has a cardinality $2^{N(H(UV_1 V_2 X_1 Z_1) +\delta)}$ and applying the joint typicality lemma \cite[Chapter 2]{el2011network}; (c) follows from the Markov chain $V_1 - U - V_2$. Following similar steps, we can prove
\begin{small}
\begin{subequations}\label{eq:IndProb}
\begin{align}
&\Pr\{\Ec_2(j)) \} \leq  2^{-N ( I(V_1;Z_2|U X_2)-\delta)},\;\;\;   \forall j\neq 1\\
&\Pr\{\Ec_3(j, k)\}   \leq 2^{-N ( I(X_1 X_2;Y|S)-\delta)}  \;\;\; \forall (j, k) \neq (1,1)\\
 &\Pr\{ \Ec_4(j, k, l, m) \}   \leq 2^{-N ( I(X_1X_2;Y|S)-\delta)},\;\;\;  \forall (j, k)\neq (1,1), \forall (l, m)\\
&\Pr\{\Ec_4(1, 1, l, m)\} \nonumber \\
 &= \Pr\{\Ec_5(l, m)\}\leq 2^{-N ( I(X_1X_2;Y|U V_1 V_2 S)-\delta)} ,\;\;\; \forall l\neq 1, m\neq 1\\
 & \Pr\{\Ec_4(1, 1, l, 1)\}\nonumber\\
 &=\Pr\{ \Ec_5(l, 1)\} \leq 2^{-N ( I(X_1;Y|U V_1 X_2 S)-\delta)},\;\;\;  \forall l\neq 1\\
  &\Pr\{\Ec_4(1, 1,1, m)\} \nonumber\\
  &= \Pr\{\Ec_5(1, m)\} \leq 2^{-N ( I(X_2;Y|U V_2 X_1 S)-\delta)},\;\;\;   \forall m\neq 1
\end{align}
\end{subequations}
\end{small}
Inserting \eqref{eq:E1} and \eqref{eq:IndProb} into \eqref{eq:UnionBound2}, we have
\begin{align}
    P_e^{(n)} \leq & (B-1) \big(2^{N \cdot R_{12}}\cdot 2^{-N \cdot I(V_1 ; Z_2|X_2 U)} \nonumber \\
    & + 2^{N \cdot R_{21}}\cdot 2^{-N \cdot I(V_2 ; Z_1|X_1 U)}\Big) \nonumber\\ 
    & + 2^{N(R_{12}+R_{21})} + 2^{-N \cdot I(X_1 X_2;Y|S)}  \nonumber \\
    & + (B-2) \Big(2^{N(R_{12}+R_{21}+R_{11}+R_{22})} \cdot 2^{-N \cdot I(X_1 X_2;Y|S)} \nonumber\\
    & + 2^{N(R_{11}+R_{22})} \cdot 2^{-N \cdot I(X_1 X_2; Y | S V_1 V_2 U)} \nonumber\\ 
    & + 2^{N \cdot R_{11}} \cdot 2^{-N \cdot I(X_1;Y|S X_2 V_1 U)} \nonumber\\
    & + 2^{N \cdot R_{22}} \cdot 2^{-N \cdot I(X_2;Y|S X_1 V_2 U)} \Big) \nonumber\\ 
    & + 2^{N(R_{11}+R_{22})} \cdot 2^{-N \cdot I(X_1 X_2;Y|S,V_1 V_2 U)} \nonumber\\
    & + 2^{N \cdot R_{11}} \cdot 2^{-N \cdot I(X_1;Y|S X_2 V_1 U)} \nonumber\\    
   &+ 2^{N \cdot R_{22}} \cdot 2^{-N \cdot I(X_2;Y|S X_1 V_2 U)} + \epsilon'_N
\end{align}
where $\epsilon'_N$ denotes a constant which vanishes as $N$ grows.
\section{Analysis of Average Distortion} \label{appendix:distortion}
For a given message pair $(w_1,w_2)$, we bound the average distortion for encoder 1.
\begin{small}
\begin{align*}
 &d_1^{(n)} (w_1,w_2) \nonumber\\
& =\EE\left[\EE\left[ \frac{1}{n} \sum_{n=1}^N d_1(S_{1i}, \hat{S}_{1i})  | X_1^n(w_1, Z^n_1), V_2^n(w_2, Z^n_2) \right]\right]\\
&\stackrel{(a)} \leq   P_e^{(n)} (w_1, w_2) d_{\max}   + (1-P_e^{(n)}(w_1, w_2) )  (1+\epsilon) \nonumber\\
& \;\;\frac{1}{n} \sum_{i=1}^n \EE\left[  \EE[d_1(S_{1i}, \upsi^*_1(x_{1}, v_{2},Z_{1i} )) | X_{1i}=x_1, V_{2i}= v_2]\right] \\
&\stackrel{(b)}= P_e^{(n)}(w_1, w_2)  d_{\max}   \nonumber\\
& \quad +(1-P_e^{(n)}(w_1, w_2) )(1+\epsilon)   \EE\left[ \uc_1(X_{1}, V_{2}) \right]. 
\end{align*}
\end{small}
where (a) follows by applying the upper bound on the
distortion function to the decoding error event and the typical
average lemma [Ch. 2.4]\cite{el2011network} to the successful decoding event; (b) follows from the cost function applied to each time $i$.  
Now, we average over all possible message pairs and obtain the average distortion for encoder 1 as:

\begin{small}
\begin{align}
d_1^{(n) } &= \frac{1}{|\Wc_1| |\Wc_2|} \sum_{(w_1, w_2)\in \Wc_1\times \Wc_2}d_1^{(n)} (w_1,w_2) \nonumber \\
&\leq \frac{(1-P_e^{(n)})(1+\epsilon) }{|\Wc_1| |\Wc_2|}  \sum_{(w_1, w_2)\in \Wc_1\times \Wc_2}\EE[ \uc_1(X_{1}(w_1), V_{2}(w_2)) ]\nonumber\\
 &+   d_{\max}  P_e^{(n)} \nonumber \\ 
& =\frac{(1-P_e^{(n)})(1+\epsilon)  }{n}  \sum_{i=1}^n \sum_{(x_1, v_2)} P_{X_{1i}, V_{2i}} (x_1, v_2) \uc_1(x_{1i}, v_{2i})\nonumber\\
 &+ d_{\max}   P_e^{(n)} 
\end{align}
\end{small}
where the last equality follows from the uniformly distributed message pair such that we have $(X_{1i}(w_1), V_{2i}(w_2))$ with probability $\frac{1}{|\Wc_1||\Wc_2|}$ for every pair $(w_1, w_2)\in \Wc_1\times \Wc_2$
Therefore, it readily follows 
\begin{align}
\limsup_{n\rightarrow\infty} d_1^{(n) } &\leq \sum_{(x_1, v_2)}P_{X_1V_2} (x_1,v_2)  \uc_1(x_{1}, v_{2}) \leq D_1.
\end{align}
Similarly, we obtain also the desired result for encoder 2.

\section{Calculation of Optimal Estimate Cost \eqref{eq:Cost}}\label{appendix:Cost}
We consider user $1$ and provide the estimator $\psi^*_1(x_1, x_2, y)$ as the solution of
\begin{align*}
\arg\min_{\psi} [P_{S_1|X_1, X_2, Y}(0| x_1, x_2, y) (0 \oplus \psi(x_1, x_2, y)) \nonumber\\
+P_{S_1|X_1, X_2, Y}(1| x_1, x_2, y)(1 \oplus \psi(x_1, x_2, y)) ]
\end{align*}
and let $d_1(x_1, x_2, y)$ denote the resulting value. Then,  the cost function is given by 
\begin{align}\label{eq:OptEstCost}
 c_1 (x_1, x_2)
=\sum_{y} P_Y(y) d_1(x_1, x_2, y)
\end{align}
We have two simple estimators. 
In the first case, the state can be estimated perfectly yielding $d_1(x_1, x_2, y)=0$. This case includes: 
\begin{align}\label{eq:226}
\psi^*_1(1,0, 0)&= 0,\;\;  \psi^*_1(1,0, 1)= 1 \nonumber \\
\psi^*_1(1,1, 0)&= 0, \;\;\psi^*_1(1,1, 2)= 1 
\end{align}
In the second case, the state of interest is erased either by the input symbol $x_1$ or the erasure event $s_1+ s_2 =1$. Then, 
we choose a fixed estimator given by 
\begin{align}\label{eq:fixedest}
\psi^*_1(x_1,x_2,y) &= 
\arg\min_{\psi\in \{0, 1\}} \left[\overline{p}_s (0 \oplus \psi) +p_s (1 \oplus \psi) \right] \nonumber \\
&= \begin{cases}
0, &  \text{if~~~} p_s < \frac{1}{2}\\
1, & \text{else}
\end{cases}
\end{align}
This fixed estimator yields $d_1(x_1, x_2, y)= \eta = \min \{p_s,1-p_s\}$ for 
\begin{align}\label{eq:241}
(x_1, x_2, y) & \in \left\{ (0,0,0), (0,1,0), (0,1,1), (1,1,1)\right\}
\end{align}
Plugging the results of \eqref{eq:226} and \eqref{eq:241} into \eqref{eq:OptEstCost}, we obtain the desired result.

\section{Calculation of Achievable Cost \eqref{eq:AchEstCost}}\label{appendix:Cost2}
We provide the estimator $\upsi^*_1(x_1, v_2, y)$ as the solution of
\begin{align*}
\arg\min_{\psi} [P_{S_1|X_1, V_2, Y}(0| x_1, v_2, y) (0 \oplus \psi(x_1, v_2, y)) \nonumber\\
+P_{S_1|X_1, V_2, Y}(1| x_1, v_2, y)(1 \oplus \psi(x_1, v_2, y)) ]
\end{align*}
and let $\ud_1(x_1, v_2, y)$ denote the resulting value. Then,  the cost function is given by 
\begin{align}\label{eq:350}
 \uc_1(x_1, v_2)=\sum_{y} P_Y(y) \ud_1(x_1, v_2, y)
\end{align}
We have three cases. 
In the first case, the state can be estimated perfectly by achieving $\ud_1(x_1, v_2, y)=0$.  We have:
\begin{align*}
\upsi^*_1(1,0, 0) &= 0, \;\;\;  \upsi^*_1(1,0,2)=1 \\ 
\upsi^*_1(1,1,0) &=  0, \;\;\; \upsi^*_1(1,1, 2)= 1
\end{align*}
In the second case, $s_1$ is erased by the associated input symbol, i.e. $x_1$ for encoder $1$. Then, 
we choose a fixed estimator independent of $(v_2, y)$ as in \eqref{eq:fixedest}. This case includes:
\begin{align}\label{eq:267}
(x_1, v_2, y) & \in \left\{ (0,0,0), (0,1,0), (0,1,1)\right\}
\end{align}
yielding $\ud_1(x_1, v_2, y)= \eta= \min \{p_s, 1-p_s\}$. In the last case, $s_1$ cannot be correctly estimated due to the interference
caused by $s_2 \theta_2$.  Noticing $P_{S_1|X_1 V_2 Y}(s_1|1,0,1)=p_s r$ if $s_1=0$ and
$P_{S_1|X_1 V_2 Y}(s_1|1,1,1)=p_s \br$ if $s_1=0$ we have
\begin{align}\label{eq:fixedest}
\upsi^*_1(1,0,1) &= 
\arg\min_{\psi\in \{0, 1\}} \left[p_s r (0 \oplus \psi) +(1-p_sr) (1 \oplus \psi) \right] \nonumber \\
&= \begin{cases}
1, &  \text{if~~~} p_s r < \frac{1}{2}\\
0, & \text{else}
\end{cases}
\end{align}
yielding $\ud_1(1,0,1)=\min \{p_s r, 1-p_s r\} = \eta_r$. Similarly we have $\ud_1(0,0,1)= \min \{p_s r, 1-p_s r\}= \eta_r$ and 
$\ud_1(1,1,1)=\min \{p_s \br, 1-p_s \br\} =\eta_{\br}$. By combining three cases and using \eqref{eq:350}, we obtain the desired result.

\section{Proof of Corollary \ref{cor:ProposedSumRate}} \label{appendix:SumRateProof}
Applying Theorem \ref{theorem:achievability} for the erasure MAC with binary states, we have
\begin{align}
R_{\rm sum-prop}(\infty) &  \leq \min\{H(Y|S), H(Y|S, V_1, V_2, U) \nonumber\\
 & + 2 \{ H(Y|X_2, U) - H(Y|X_2, U, V_1) \}\} 
 \end{align}
In order to characterize each term inside $\min$, we first provide the input and output distribution. We have
 \begin{align}
 P_{X_1, X_2} (0,0) 
 &= \bp \kappa^2 + p\bkappa^2 , \;\;\;\;  P_{X_1, X_2} (1,1)=p \kappa^2 +\bp \bkappa^2\nonumber\\
 P_{X_1, X_2} (0,1) 
 &=  P_{X_1, X_2} (1,0) =\kappa \bkappa 
 \end{align}
 as well as
  \begin{align}
P_Y(0) 
&= P_{X_1, X_2} (0, 0)+ 2 \bPs P_{X_1, X_2}(0,1) + \bPs^2 P_{X_1, X_2}(1,1) \nonumber\\
P_Y(1) 
&= 2 p_s P_{X_1, X_2} (0, 1) + 2 p_s \bPs P_{X_1, X_2} (1, 1)\nonumber\\
P_Y(2) 
&=p_s^2 P_{X_1, X_2} (1, 1) 
\end{align}

Now we will examine each term inside $\min$. \\

{\bf  Term $f_2(p, q, r)=  H(Y|S) $}\\
\begin{small}
\begin{align}\label{eq:411}
f_1 (p, q, r) 
&=- \sum_{(s_1, s_2)} P_{S_1}(s_1) P_{S_2}(s_2)  \sum_{y} P_{Y|S_1 S_2}(y|s_1, s_2)\\
& \;\;  \log  P_{Y|S_1 S_2}(y|s_1, s_2)\\
\end{align}
\end{small}
where 
\begin{small}
\begin{align*}
P_{Y|S_1 S_2}(0|0,0) &= 1 , P_{Y|S_1 S_2}(y|0,0)=0, \forall y \in \{1, 2\}\\
P_{Y|S_1 S_2}(0|0,1) &= P_{X_2}(0) = \bp \kappa^2 + p \bkappa^2 +\kappa \bkappa\\
P_{Y|S_1 S_2}(1|0,1) &= P_{X_2}(1) = \kappa \bkappa + p \kappa^2 +\bp \bkappa^2 \\
 P_{Y|S_1 S_2}(0|1,1) &=P_{X_1X_2}(0,0) =\bp \kappa^2 + p\bkappa^2 \\
 P_{Y|S_1 S_2}(1|1,1) &=P_{X_1X_2}(0,1)+P_{X_1X_2}(1,0)= 2 \kappa \bkappa\\
P_{Y|S_1 S_2}(2|1,1) &=P_{X_1X_2}(1,1) = p \kappa^2 +\bp \bkappa^2 \\
\end{align*}
\end{small}
Plugging these expressions into \eqref{eq:411}, we obtain
\[ f_2 =2 p_s\bPs H_2(p\kappa+ \bp\bkappa) + p_s^2 H_3(\bp \kappa^2 + p\bkappa^2 , 2\kappa \bkappa, p \kappa^2 +\bp \bkappa^2 )\]

{\bf Term $f_{1a}=H(Y|S, V_1, V_2, U) $}\\
By defining 
\[ Y'= S_1\Theta_1 + S_2 \Theta_2 \]
we have
\begin{align}
&H(Y|S, V_1, V_2, U) = H(Y'|S) \nonumber\\
&= \sum_{(s_1 s_2)} P_{S_1S_2}(s_1, s_2) \sum_{y'} P_{Y'|S} (y'|s) \log P_{Y'|S} (y'|s) 
\end{align}
where we have
\begin{align*}
P_{Y'|S} (y'|00) & = \onev\{y'=0\} \\
P_{Y'|S} (y'|01) & = \br \onev\{y'=0\}  +  r \onev\{y'=1\}  \\
P_{Y'|S} (y'|01) & = \br \onev\{y'=0\}  +  r \onev\{y'=1\}  \\
P_{Y'|S} (y'|11) & = \br^2 \onev\{y'=0\}  +  2 r\br \onev\{y'=1\}  + r^2 \onev\{y'=2\}  
\end{align*}
yielding
\begin{align*}
f_{1a} &= 2  \bPs p_s H_2(r)  + p_s^2H_3(r^2, 2r \br, \br^2).
\end{align*}

{\bf Term $f_{1b}=H(Y|X_2, U) $}
\begin{align}\label{eq:743}
H(Y|X_2, U) &= \sum_u P_U(u) \sum_{x_2}  P_{X_2|U} (x_2|u) \nonumber\\
& \quad\sum_y P_{Y|X_2 U} (y|x_2 u)  \log P_{Y|X_2 U}(y|x_2 u)
\end{align}
with 
\begin{align}\label{eq:749}
 P_{Y|X_2, U}(y|x_2, u) &= \sum_{s} P_S(s) P_{Y|S,X_2, U}(y|s,x_2, u) \nonumber\\
&= \bPs^2 \onev\{y =0, \forall (x_2, u) \} \nonumber\\
&+ \bPs p_s \onev\{y = x_2, \forall u) \} \nonumber\\
& + p_s\bPs P_{X_1|U}(y|u) \onev\{y\in \{0, 1\} \} \nonumber\\
&+ p_s^2 P_{X_1|U}(y-x_2|u) \onev\{(y-x_2)\in \{0, 1\} \}
\end{align}
yielding 
\begin{small}
\begin{align}\label{eq:320}
 P_{Y|X_2, U}(0|0,0) &=  \bPs + p_s \kappa \nonumber\\ 
 P_{Y|X_2, U}(1|0,0)  &=p_s \bkappa \nonumber\\
 P_{Y|X_2, U}(2|0,0) & = 0  \nonumber\\ 
 P_{Y|X_2, U}(0|0,1) & = \bPs + p_s \bkappa \nonumber\\ 
 P_{Y|X_2, U}(1|0,1)  & =p_s \kappa \nonumber\\
 P_{Y|X_2, U}(2|0,1) & = 0  \nonumber\\ 
 P_{Y|X_2, U}(0|1,0) & =\bPs^2 + p_s\bPs \kappa\nonumber\\
 P_{Y|X_2, U}(1|1,0)   &= \bPs p_s +  p_s\bPs \bkappa + p_s^2 \kappa \nonumber \\
 P_{Y|X_2, U}(2|1,0) & = p_s^2 \bkappa \nonumber \\ 
  P_{Y|X_2, U}(0|1,1) & =\bPs^2+ p_s\bPs \bkappa \nonumber\\
  P_{Y|X_2, U}(1|1,1)  &= \bPs p_s+  p_s\bPs \kappa + p_s^2 \bkappa \nonumber \\
  P_{Y|X_2, U}(2|1,1) &= p_s^2 \kappa 
\end{align}
\end{small}
where we used
\begin{subequations}
\begin{align}\label{eq:754}
P_{X_1|U} (0|0)=P_{X_1|U} (1|1) =  \kappa \\
P_{X_1|U} (0|1)=P_{X_1|U} (1|0) =  \bkappa
\end{align}
\end{subequations}
Plugging  \eqref{eq:320} into \eqref{eq:743} and using
\begin{subequations}
\begin{align}
P_{X_2 U}(0,0) &= P_U(0) P_{X_2|U}(0|0) = \bp \kappa\\
P_{X_2 U}(0,1) &= P_U(1) P_{X_2|U}(0|1) = p \bkappa \\
P_{X_2 U}(1,0) &= P_U(0) P_{X_2|U}(1|0) = \bp \bkappa\\
P_{X_2 U}(1,1) &= P_U(1) P_{X_2|U}(1|1) = p \kappa 
\end{align}
\end{subequations}
 we obtain the desired expression for $f_{1b}$. 
 
%

{\bf Term : $f_{1c}=H(Y|U, V_1,X_2)$}\\

We have
\begin{small}
\begin{align}\label{eq:526}
&H(Y|U,V_1,X_2) = \sum_u P_U(u) \sum_{x_2}  P_{X_2|U} (x_2|u) \sum_{v_1} P_{V_1|U} (v_1|u) \nonumber\\
& \sum_y P_{Y|U V_1X_2} (y|u v_1 x_2)  \log P_{Y|U V_1 X_2}(y| u v_1 x_2)
\end{align}
\end{small}
where 
\begin{align*}
& P_{Y|U, V_1,X_2} (y|u, v_1,x_2)   =  P_{Y|V_1,X_2} (y|v_1,x_2) \\
 & =\sum_{s} P_S(s) P_{Y|S, V_1, X_2} (y|s, v_1,x_2) \\
 & = \bPs^2 \onev\{y =0, \forall (x_2 v_1)\}+ p_s \bPs \onev\{ y =x_2, \forall v_1\}\\
&+  p_s\bPs P_{\Theta_1}(y\oplus v_1) \onev\{ y\in \{0,1\} ,\forall x_2\} \nonumber\\
&+ p_s^2 P_{\Theta_1}(( y-x_2 )\oplus v_1) \onev\{(y-x_2)\in \{0,1\}\}
\end{align*}
yielding 
\begin{align*}
 P_{Y|V_1,X_2} (0|0,0)  &= \bPs+  p_s  \br \\
 P_{Y|V_1,X_2} (1|0,0)  & =p_s r \\
 P_{Y|V_1,X_2} (2|0,0)  &=0 \\
  P_{Y|V_1,X_2} (0|0,1)  &=  \bPs^2 + p_s\bPs\br \\
    P_{Y|V_1,X_2} (1|0,1) &=  p_s \bPs + p_s\bPs r + p_s^2  \br \\
    P_{Y|V_1,X_2} (2|0,1)  &=p_s^2 r\\
    \end{align*}
   
    \begin{align*}
 P_{Y|V_1,X_2} (0|1,0)  &= \bPs +p_s r \\
 P_{Y|V_1,X_2} (1|1,0) &= p_s \br \\
   P_{Y|V_1,X_2} (2|1,0)&= 0\\
    P_{Y|V_1,X_2} (0|1,1)  &=    \bPs^2+p_s\bPs r\\ 
     P_{Y|V_1,X_2} (1|1,1)&=p_s \bPs+p_s\bPs  \br+  p_s^2 r\\
      P_{Y|V_1,X_2} (2|1,1)&=p_s^2 \br
\end{align*}
By noticing
\begin{align*}
P_{V_1X_2}(v_1, x_2) &= \sum_u  P_U(u) P_{V_1|U} (v_1|u) P_{X_2|U} (x_2|u)\\
& = \bp P_{\Sigma_1} (v_1)  P_{X_2|U} (x_2|0) + p P_{\Sigma_1} (\bar{v}_1) P_{X_2|U} (x_2|1)
\end{align*}
we readily obtain
\begin{align*}
P_{V_1X_2}(0,0)&= \bp \bq \kappa + p q \bkappa\\
P_{V_1X_2}(0,1)&= \bp \bq  \bkappa + p q \kappa\\
P_{V_1X_2}(1,0)&= \bp q \kappa+ p \bq \bkappa \\
P_{V_1X_2}(1,1)&= \bp  q \bkappa + p \bq \kappa 
\end{align*}
Plugging these expressions into \eqref{eq:526}, we obtain the desired result.

\end{document}